\documentstyle[12pt,preprint,aps,tighten,epsfig,floats]{revtex}

\def\beq{\begin{equation}}
\def\eeq{\end{equation}}
\def\beqa{\begin{eqnarray}}
\def\eeqa{\end{eqnarray}}
\def\be{\begin{equation}}
\def\ee{\end{equation}}
\def\bea{\begin{eqnarray}}
\def\eea{\end{eqnarray}}
\begin{document}

\title{Dispersive calculation of $B_7^{(3/2)}$ and $B_8^{(3/2)}$ in
the chiral limit}
\author{John F. Donoghue$^{1,2}$ and Eugene Golowich$^1$} 
\address{$^1$Department of Physics and Astronomy, 
University of Massachusetts\\
Amherst, Massachusetts 01003  \\ and \\ $^2$TH Division, CERN, Geneva}
\maketitle
\thispagestyle{empty}
\setcounter{page}{0}
\begin{abstract}

\noindent We show how the isospin vector and axialvector 
current spectral functions $\rho_{V,3}$ and $\rho_{A,3}$ can be 
used to determine in leading chiral order the low energy constants 
$B_7^{(3/2)}$ and $B_8^{(3/2)}$. This is accomplished by matching the
Operator Product Expansion to the dispersive analysis of vacuum
polarization functions. The data for the evaluation of these
dispersive integrals has been recently enhanced by the ALEPH
measurement of spectral functions in tau decay, and we update our
previous phenomenological determination. Our 
calculation yields in the NDR renormalization scheme 
and at renormalization scale $\mu = 2~{\rm GeV}$ the values 
$B_7^{(3/2)} = 0.55 \pm 0.07 \pm 0.10$ and 
$B_8^{(3/2)} = 1.11 \pm 0.16 \pm 0.23$ for the quark mass values 
$m_s + {\hat m} = 0.1$~GeV.
\end{abstract}

\vspace{1.0in}

\section{Introduction}
The recent KTeV and NA48 findings that $\epsilon'/\epsilon 
\simeq 20 \cdot 10^{-4}$ raise the important question 
whether a value so large can be consistent with Standard Model 
expectations.~\cite{ktev}  One of the key quantities upon which the 
Standard Model prediction is based is the B-factor $B_8^{(3/2)}$. 
In this paper, we work in the chiral limit to obtain an 
analytic expression for $B_8^{(3/2)}$ (and also for $B_7^{(3/2)}$).  
Our results take the form of sum rules involving the difference 
$\rho_V - \rho_A$ of isospin-one spectral functions.

The constants $B_7^{(3/2)}(\mu)$ and $B_8^{(3/2)}(\mu)$ are 
defined in terms of the matrix elements 
\beq
\langle 2\pi | {\cal Q}_i^{(3/2)} | K \rangle_\mu \equiv 
B_i^{(3/2)}(\mu) ~
\langle 2\pi | {\cal Q}_i^{(3/2)} | K \rangle_\mu^{\rm vac} 
\qquad ( i = 7,8) \ \ ,
\label{b2}
\eeq
where $\mu$ is the renormalization scale, 
${\cal Q}_7^{(3/2)}$ and ${\cal Q}_8^{(3/2)}$ are 
the $\Delta I = 3/2$ electroweak penguin operators 
\beqa
{\cal Q}_7^{(3/2)} &\equiv& {\bar s}_a\Gamma^\mu_{\rm L}d_a 
\left( {\bar u}_b\Gamma_\mu^{\rm R}u_b - 
{\bar d}_b\Gamma_\mu^{\rm R}d_b \right) + 
{\bar s}_a\Gamma^\mu_{\rm L}u_a {\bar u}_b\Gamma_\mu^{\rm R}d_b  
\ \ ,
\nonumber \\
{\cal Q}_8^{(3/2)} &\equiv& {\bar s}_a\Gamma^\mu_{\rm L}d_b
\left( {\bar u}_b\Gamma_\mu^{\rm R}u_a - 
{\bar d}_b\Gamma_\mu^{\rm R}d_a \right) + 
{\bar s}_a \Gamma^\mu_{\rm L}u_b {\bar u}_b\Gamma_\mu^{\rm R}d_a 
\ \ ,
\label{b1a}
\eeqa
$a,b$ are color labels and $\Gamma^\mu_{\rm L} \equiv 
\gamma^\mu (1 + \gamma_5)$, $\Gamma^\mu_{\rm R} \equiv 
\gamma^\mu (1 - \gamma_5)$. Analogous B-factors are defined for 
the other weak operators. The most important contributions to 
$\epsilon'/\epsilon$ are the matrix elements of the 
penguin operator ${\cal Q}_6$ and the
electroweak penguin operator ${\cal Q}^{3/2}_8$. 
As expressed in terms of the B-factors
this is seen in the approximate numerical relation, 
\beq
{\epsilon' \over \epsilon} = 10 \times 10^{-4} \left[ 
(2.5 B_6 - 1.3 B_8^{(3/2)})
\left( {100~{\rm MeV} \over m_s + m_d}\right) \right] \ \ ,
\eeq
when evaluated at $\mu=2$~GeV in the ${\overline {MS}}$-NDR 
renormalization scheme.  Alternatively, in terms of the operator 
matrix elements themselves 
one has
\beq
{\epsilon' \over \epsilon} = 10 \times 10^{-4} \left[ 
-3.1~{\rm GeV}^{-3} \cdot \langle {\cal Q}_6\rangle_0 
- 0.51~{\rm GeV}^{-3} \cdot \langle{\cal Q}_8^{(3/2)}\rangle_2
\right]
\eeq
where
\beq
\langle {\cal Q}_6\rangle_0 \equiv \langle (\pi\pi)_{I=0} | 
{\cal Q}_6 | K^0\rangle \qquad {\rm and} \qquad 
\langle {\cal Q}_8^{(3/2)}\rangle_2 \equiv
\langle (\pi\pi)_{I=2} | {\cal Q}_8^{(3/2)} | K^0\rangle \ \ ,
\eeq
again defined in the NDR scheme at $\mu = 2$~GeV.

A kaon-to-pion weak matrix element can be analyzed by chiral 
methods and expressed as an expansion in momenta and quark 
masses.~\cite{dgh}  Here we shall calculate the leading term 
in such an expansion, valid in the limit of exact chiral 
symmetry where the Goldstone bosons become massless.  In fact, 
our analysis yields values for the relevant $K$-to-$\pi$ 
matrix elements themselves.  However, we shall transcribe this 
information into the equivalent form of B-factors 
in order to express our results in the more conventional form 
and thus allow comparison with other techniques. When considered 
in the chiral limit, the content of Eq.~(\ref{b2}) reduces to 
\beqa
\lim_{p=0}~ B_7^{(3/2)}(\mu) &=& -{3 \over F_\pi^4} \cdot
{m_u + m_d \over m_\pi^2} \cdot {m_u + m_s \over m_K^2} 
~\langle {\cal O}_1 \rangle_\mu \ \ ,
\nonumber \\
\lim_{p=0}~ B_8^{(3/2)} (\mu) &=& -{1 \over F_\pi^4} \cdot
{m_u + m_d \over m_\pi^2} \cdot {m_u + m_s \over m_K^2} ~\left[ 
{1 \over 3} \langle {\cal O}_1 \rangle_\mu + 
{1 \over 2} \langle {\cal O}_8 \rangle_\mu \right] \ \ , 
\label{lat10} 
\eeqa
where ${\cal O}_1$, ${\cal O}_8$ are the local four-quark operators 
\beqa
{\cal O}_1 &\equiv& {\bar q} \gamma_\mu {\tau_3 \over 2} q
~{\bar q} \gamma^\mu {\tau_3 \over 2} q - 
{\bar q} \gamma_\mu \gamma_5 {\tau_3 \over 2} q 
~{\bar q} \gamma^\mu \gamma_5 {\tau_3 \over 2} q \ \ ,
\nonumber \\
{\cal O}_8 &\equiv& {\bar q} \gamma_\mu \lambda^A
{\tau_3 \over 2} q
~{\bar q} \gamma^\mu \lambda^A {\tau_3 \over 2} q - 
{\bar q} \gamma_\mu \gamma_5 \lambda^A {\tau_3 \over 2} q
~{\bar q} \gamma^\mu \gamma_5 \lambda^A {\tau_3 \over 2} q
\ \ .
\label{sd4}
\eeqa
In the above, $q = u,d,s$, $\tau_3$ is a Pauli (flavor) matrix, 
$\{ \lambda^A \}$ are the Gell~Mann color 
matrices and the subscripts on ${\cal O}_1$, ${\cal O}_8$ 
refer to the color carried by their currents.\footnote{Throughout we 
denote vacuum expectation values as $\langle 0 | {\cal O} | 0 
\rangle_\mu \equiv \langle {\cal O} \rangle_\mu $.}
A chiral evaluation of $B_7^{(3/2)}$ and $B_8^{(3/2)}$ is 
thus seen ({\it cf} Eq.~(\ref{lat10})) to depend upon 
$\langle {\cal O}_1 \rangle_\mu$ and 
$\langle {\cal O}_8 \rangle_\mu$. 
The $K^0 \to \pi\pi$ matrix elements are likewise recoverable 
in the chiral limit from the vacuum matrix elements, 
\beqa
\lim_{p=0}~ \langle (\pi\pi)_{I=2} | {\cal Q}_7^{(3/2)} 
|K^0\rangle_\mu & = & - {4 \over F_\pi^3}~\langle {\cal O}_1 
\rangle_\mu  \ \ , \nonumber \\ 
\lim_{p=0}~ \langle (\pi\pi)_{I=2} | {\cal Q}_8^{(3/2)} | 
K^0\rangle_\mu & = & - {4 \over F_\pi^3} ~\left[ 
{1 \over 3} \langle {\cal O}_1 \rangle_\mu + 
{1 \over 2} \langle {\cal O}_8 \rangle_\mu \right] \ \ .
\label{b7b8}
\eeqa 
The rest of this paper describes a calculational procedure to 
obtain analytic expressions for the vacuum matrix elements.  

In Section II, we show how to extract 
$\langle {\cal O}_1 \rangle_\mu$ and 
$\langle {\cal O}_8 \rangle_\mu$ from 
the isospin-one vector and axialvector correlators by 
deriving sum rules for $\langle {\cal O}_1 \rangle_\mu$ and 
$\langle {\cal O}_8 \rangle_\mu$ in a momentum cutoff scheme.  
This renormalization is especially well suited for comparing 
theory to experiment.  In Section III we introduce a 
four-quark operator ${\cal O}_{\Delta S = 1}$, 
distinct from the familiar nonleptonic hamiltonian 
${\cal H}_{\Delta S = 1}$, whose kaon-to-pion matrix element is 
related to $\langle {\cal O}_1 \rangle_\mu$ and $\langle 
{\cal O}_8 \rangle_\mu$ in the chiral limit.  
We demonstrate consistency of this information with the 
sum rules of Section II.  Section IV describes a procedure for 
obtaining $\langle {\cal O}_1 \rangle_\mu$ and $\langle 
{\cal O}_8 \rangle_\mu$ in 
${\overline {MS}}$ renormalization, which is commonly used 
in lattice-theoretic simulations.  Our final numerical results 
and concluding statements appear respectively in Sections V and VI. 

\section{Dispersive Analysis of Vacuum Polarization Functions}
In seeking values for $\langle {\cal O}_1 \rangle_\mu$ and $\langle 
{\cal O}_8 \rangle_\mu$, it is natural to study the 
vacuum polarization functions as these are also defined 
in terms of vacuum matrix elements of four-quark 
operators (but generally not all at the same spacetime point). 
Thus we consider the combination 
$\Pi_{V,3} - \Pi_{A,3}$ (the subscript `$3$' denotes 
the isospin flavor),  
\beqa
& & i \int d^4 x\ e^{i q \cdot x} 
\langle 0 |T\left( V^\mu_3  (x) V^\nu_3 (0) - 
A^\mu_3 (x) A^\nu_3 (0)\right) | 0 \rangle 
\nonumber \\
& & \phantom{xxxxx} = (q^\mu q^\nu - q^2 g^{\mu\nu} ) 
( \Pi_{V,3} - \Pi_{A,3} )(q^2) - q^\mu q^\nu \Pi_{A,3}^{(0)}(q^2) 
\ \ .
\label{r0}
\eeqa
Associated with this correlator is the 
difference of spectral functions $\rho_{V,3} - \rho_{A,3}$, 
\beq
\left[ \Pi_{V,3} - \Pi_{A,3} \right](Q^2) = {1 \over Q^4} 
\int_0^\infty ds\  {s^2 \over s + Q^2}~\left[ 
\rho_{V,3} - \rho_{A,3} \right](s) \ \ ,
\label{sp1}
\eeq
where $Q^2 \equiv - q^2$. In writing this spectral relation, 
we have made use of the first and second Weinberg sum 
rules~\cite{sw}, which are both valid in the chiral limit

Due to the complexity of QCD, there exist no analytic 
expressions for the correlators and spectral functions 
that are valid over the entire energy domain.  However, 
some crucial information is available. 
At low energies, $\rho_{V,3}$ and $\rho_{A,3}$ are determined 
from $\tau$-lepton decays and from $e^+e^-$ scattering. 
As one proceeds from the resonance region of nonperturbative 
physics to larger energies, the effect of individual channels 
becomes indistinguishable and perturbative QCD (pQCD) 
becomes operative.  The boundary between nonperturbative 
and perturbative regions defines a scale $\Lambda \sim 2\to 3$~GeV.  
In the pQCD domain, the leading-log behavior of 
$(\Pi_{V,3} - \Pi_{A,3})(Q^2)$ is given by~\cite{lanin}
\beq
Q^6 (\Pi_{V,3} - \Pi_{A,3})(Q^2) \sim 
2 \pi \langle \alpha_s {\cal O}_8 \rangle_{\mu} + 
\ln \left( {Q^2 \over \mu^2} 
\right) \left[ {8 \over 3} \langle \alpha_s^2 {\cal O}_1 
\rangle_{\mu} - \langle \alpha_s^2 {\cal O}_8 \rangle_{\mu}  
\right] + \dots \ \ .
\label{c2}
\eeq
This asymptotic relation will be of special value 
to our determination of $B_8^{(3/2)}$ 
since it contains information on the vacuum matrix element 
$\langle {\cal O}_8 \rangle_{\mu}$ ({\it cf} Eq.~(\ref{b7b8})). 
The large-$s$ behavior of the spectral functions can be 
inferred from the logarithmic term in Eq.~(\ref{c2}) via 
continuation to the real $q^2$-axis, 
\beq
(\rho_{V,3} - \rho_{A,3})(s) \sim {1 \over s^3} 
\left[ {8 \over 3} \langle \alpha_s^2 {\cal O}_1 \rangle_\mu
-  \langle \alpha_s^2 {\cal O}_8 \rangle_\mu \right] + \dots 
\ \ .
\label{sp2}
\eeq
Together, the spectral relations of Eq.~(\ref{sp1}) and Eq.~(\ref{sp2}) 
imply 
\beqa
& & Q^6 (\Pi_{V,3} - \Pi_{A,3})(Q^2) \ \sim \ \ln 
\left( {Q^2 \over \Lambda^2} \right) ~
\left[ {8 \over 3} \langle \alpha_s^2 {\cal O}_1 
\rangle_{\mu} - \langle \alpha_s^2 {\cal O}_8 \rangle_{\mu}  
\right] 
\nonumber \\
& & \phantom{xxxxxxxxxx} + \int_0^{\Lambda^2} ds\  s^2 ~\left[ 
\rho_{V,3} - \rho_{A,3} \right](s) + {\cal O}(Q^{-2}) \ \ .
\label{sp2p}
\eeqa

\subsection{Correlators in $d$-Dimensions}
Consider the definition of $\Pi_{V,3} - \Pi_{A,3}$ 
as expressed in $d$-dimensions, 
\beqa
& & \mu_{\rm d.r.}^{d-4} i \int d^dx\ e^{i q \cdot x} 
\langle 0 |T\left( V^\mu_3  (x) V^\nu_3 (0) - 
A^\mu_3 (x) A^\nu_3 (0)\right) | 0 \rangle 
\nonumber \\
& & \phantom{xxxxx} = (q^\mu q^\nu - q^2 g^{\mu\nu} ) 
( \Pi_{V,3} - \Pi_{A,3} )(q^2) - q^\mu q^\nu \Pi_{A,3}^{(0)}(q^2) \ \ .
\label{r1}
\eeqa
The energy scale $\mu_{\rm d.r.}$ (`d.r.' denotes dimensional 
regularization) has been introduced to maintain 
the proper dimensions away from $d=4$.  It 
is straightforward to invert Eq.~(\ref{r1}) and we find
\beqa
& &  \langle 0 |T\left( V^\mu_3  (x) V_{\mu,3}  (0) - 
A^\mu_3 (x) A_{\mu, 3} (0)\right) | 0 \rangle 
\nonumber \\
& & \phantom{xxxxx} = { (d - 1) \mu_{\rm d.r.}^{4 -d} 
\over (4 \pi)^{d/2} \Gamma(d/2)} \int_0^\infty dQ^2 
\ e^{-i q \cdot x}~Q^d  \left( \Pi_{V,3} - \Pi_{A,3} \right)(Q^2) \ \ .
\label{r2}
\eeqa
Up to this point the procedure is well defined, as all 
quantities are finite-valued.  

To obtain a relation for $\langle {\cal O}_1 \rangle_\mu$, we 
need to evaluate Eq.~(\ref{r2}) in the limit of $x \to 0$.  
However, the asymptotic condition of Eq.~(\ref{c2}) implies that 
unless the integral on the right-hand-side of Eq.~(\ref{r2}) 
is regularized, it will diverge as $x \to 0$.  
There are a number of ways to perform the regularization, 
and we shall consider two particularly useful approaches --- 
first a momentum space cutoff directly below and then 
${\overline {MS}}$ renormalization in Sect.~IV.  We shall 
distinguish vacuum matrix elements in the two schemes by means 
of the superscripts `(c.o.)' for momentum-cutoff 
and `(${\overline {MS}}$)' for modified minimal subtraction.  

\subsection{Two Sum Rules in Momentum-space Cutoff Renormalization}
Let us remove the divergence which occurs for $d=4$ in 
Eq.~(\ref{r2}) by cutting off the $Q^2$-integral 
at $Q^2 = \mu^2$, where $\mu$ is the renormalization 
scale and for convenience we set $\mu_{\rm d.r.} = \mu$.  It is 
valid to take $d=4$ in this case since the integral is finite. 
We find
\beq
\langle {\cal O}_1 \rangle_{\mu}^{\rm (c.o.)} 
= {3 \over 16 \pi^2} 
\int_0^{\mu^2} dQ^2 \ Q^4 ( \Pi_{V,3} - \Pi_{A,3} )(Q^2) \ \ .
\label{r3}
\eeq
Using Eq.~(\ref{sp1}) to express this relation in terms of 
spectral functions, we arrive immediately at the following 
sum rule, 
\beq
{16 \pi^2 \over 3} \langle {\cal O}_1 \rangle_{\mu}^{\rm (c.o.)} 
= I_1 \equiv 
\int_0^\infty  ds\  s^2 \ln \left({s + \mu^2 \over s} \right)
~\left[ \rho_{V,3} - \rho_{A,3} \right](s)  \ \ .
\label{sp5} 
\eeq

It is equally straightforward to derive a sum rule for 
$\langle \alpha_s {\cal O}_8 \rangle_{\mu}^{\rm (c.o.)}$. 
We first set $Q^2 = \mu^2$ in Eq.~(\ref{c2}) to obtain 
\beq
\langle \alpha_s {\cal O}_8 \rangle_{\mu}^{\rm (c.o.)} = 
{ \mu^6 \over 2 \pi} (\Pi_{V,3} - \Pi_{A,3})(\mu^2) \ \ .
\label{sp2aa}
\eeq
Because the variable $Q^2$ is constrained 
in Eq.~(\ref{c2}) to lie in the range where pQCD makes sense, 
the same must be true for the scale $\mu$.  Then by combining 
Eq.~(\ref{sp2aa}) with Eq.~(\ref{sp1}), we obtain the sum rule 
\beq
2 \pi \langle \alpha_s {\cal O}_8 \rangle_{\mu}^{\rm (c.o.)} 
= I_8 \equiv \int_0^\infty ds\  s^2 {\mu^2 \over s + \mu^2}
~\left[ \rho_{V,3} - \rho_{A,3} \right](s) \ \ .
\label{sr31}
\eeq

Despite their apparent similarity, it is important to 
understand that there 
is a basic difference between the sum rules for 
$\langle {\cal O}_1 \rangle_{\mu}^{\rm (c.o.)}$ 
and $\langle \alpha_s {\cal O}_8 \rangle_{\mu}^{\rm (c.o.)}$.  
The former is obtained rather directly by taking the $x \to 0$ 
limit of Eq.~(\ref{r2}) and using a cutoff in momentum to 
regularize the procedure.  However, the latter rests upon 
assuming the dominance of the leading $Q^{-6}$ term in the 
OPE of Eq.~(\ref{c2}).  This assumption becomes increasingly 
questionable as $\mu$ is lowered to energies just above the 
resonance region.  It leads to an uncertainty in the value of 
$\langle \alpha_s {\cal O}_8 \rangle_{\mu}^{\rm (c.o.)}$ 
which is not present in 
$\langle {\cal O}_1 \rangle_{\mu}^{\rm (c.o.)}$.  
We postpone discussion to Sect.~V regarding numerical evaluation 
of the integrals $I_1$, $I_8$ appearing in 
Eqs.~(\ref{sp5}),(\ref{sr31}).

\section{Kaon-to-pion Matrix Elements of a Left-Right Operator}
A distinct but equivalent path to learn about 
$\langle {\cal O}_1 \rangle_\mu$ and $\langle 
{\cal O}_8 \rangle_\mu$ is to perform a chiral analysis of 
the kaon-to-pion matrix elements themselves.  However, the 
usual (V-A)$\times$(V-A) weak hamiltonian 
${\cal H}_{\Delta S = 1}$ would be of no help 
in the chiral limit since its K-to-pi matrix elements vanish 
there.  Instead we introduce a (V-A)$\times$(V+A) 
nonleptonic operator ${\cal O}_{\Delta S = 1}$ defined as 
\beqa
{\cal O}_{\Delta S = 1}  &\equiv& 
{g_2^2 \over 8} \int d^4x \ 
{\cal D}_{\mu\nu}(x,M_W^2) ~J^{\mu\nu}(x) \ \ ,
\nonumber \\
J^{\mu\nu}(x) &\equiv& {1\over 2} 
T\left[ {\bar d}(x) \gamma^\mu (1 + \gamma_5 ) u (x) ~
{\bar u}(0) \gamma^\nu (1 - \gamma_5 ) s (0) \right]
\nonumber \\
&=& {1\over 2} 
T\left[ \left(V^\mu_{1 - i2} (x) + A^\mu_{1 - i2} (x)\right) ~
\left(V^\nu_{4 + i5} (0) - A^\nu_{4 + i5} (0) \right)
\right] \ \ ,
\label{va}
\eeqa
where ${\cal D}_{\mu\nu}$ is the $W$-boson propagator and $V^\mu_a$, 
$A^\mu_a \ (a = 1,\dots 8)$ are the flavor-octet vector, 
axialvector currents.  Operators 
similar to ${\cal O}_{\Delta S = 1}$ have received some previous 
attention in the literature.~\cite{dg1,davier}  The LR chiral
structure of ${\cal O}_{\Delta S = 1}$ ensures the survival of 
the K-to-pi matrix element ${\cal M}(p) =  \langle \pi^- (p)| 
{\cal O}_{\Delta S = 1} | K^- (p)\rangle$ in the $p \to 0$ limit,  
where we obtain 
\beq
{\cal M} \equiv \lim_{p=0}{\cal M}(p) 
= {g_2^2 \over 16 F_\pi^2} \int d^4x \ {\cal D}(x,M_W^2) ~
\langle 0 |T\left( V^\mu_3  (x) V_{\mu,3}  (0) - 
A^\mu_3 (x) A_{\mu, 3} (0)\right) | 0 \rangle \ \ .
\label{va3}
\eeq 

\subsection{Leading-log Analysis of QCD Corrections}
In the following, we perform a leading-log calculation 
of QCD corrections to the chiral matrix element ${\cal M}$. 
This leads naturally to renormalization group equations (RGE) for 
the quantities $\langle {\cal O}_1 \rangle_\mu$ and $\langle \alpha_s 
{\cal O}_8 \rangle_\mu$. 

Since the $W$-boson propagator in Eq.~(\ref{va3}) acts as a cutoff 
for contributions with $|x| \ge M_W^{-1}$, we consider the 
leading term of the following operator product expansion, 
\beq
V^\mu_3  (x) V^\nu_3  (0) - 
A^\mu_3 (x) A^\nu_3 (0) = 
V^\mu_3  (0) V^\nu_3  (0) - 
A^\mu_3 (0) A^\nu_3 (0) + {\cal O}(x) \ \ .
\label{sd1}
\eeq
Evaluation of the spacetime integral in Eq.~(\ref{va3}) is 
straightforward, 
\beq
\int d^4x \ {\cal D}_{\mu\nu}(x,M_W^2) = 
{g_{\mu\nu} \over M_W^2} \ \ , 
\label{sd2}
\eeq
so that the matrix element specified at energy scale 
$M_W$ becomes 
\beq
{\cal M} \simeq {G_F \over 2 \sqrt{2}F_\pi^2} 
\langle {\cal O}_1 \rangle_{M_W}\ \ .
\label{sd3}
\eeq 

In order to express this vacuum matrix element at some lower energy 
$\mu$, we must take QCD radiative corrections into account.  
The effect of these will be to mix ${\cal O}_1$ with ${\cal O}_8$.  
The result of mixing at one-loop order is 
\beq
{{\cal O}_1 \brack {\cal O}_8 }
\ \to\   {{\cal O}_1 \brack {\cal O}_8 }
\ + \ { \alpha_s \over 4 \pi}\ln \left( {M_W^2 \over \mu^2} \right) 
\left[ \begin{array}{ll} 
\ \  0 & 3/2 \\ 
16/3 & ~\ 7 
\end{array}\right] 
{{\cal O}_1 \brack {\cal O}_8 } \ \ ,
\label{sd6}
\eeq
where $\alpha_s \equiv g_3^2 /4 \pi$ is the QCD fine structure 
constant and $\mu < M_W$.  Using standard 
techniques~\cite{dgh}, we use the renormalization group (RG) 
to provide a summation of the leading-log dependence 
over the range from $M_W$ down to $\mu$, 
\beq
{\cal M} \simeq {G_F \over 2 \sqrt{2}F_\pi^2} 
\left[ c_1 (\mu) \langle {\cal O}_1 \rangle_{\mu} 
+ c_8 (\mu) \langle {\cal O}_8 \rangle_{\mu}  
\right] \ \ ,
\label{sd7}
\eeq
where 
\beqa
c_1 (\mu) &=& {1 \over 9} \left[ \left( {\alpha_s (\mu) 
\over \alpha_s (M_W)} \right)^{8/9} + 8 
\left( {\alpha_s (\mu) \over \alpha_s (M_W)} 
\right)^{-1/9} \right] \ \ , 
\nonumber \\
c_8 (\mu) &=& {1 \over 6} \left[ \left( {\alpha_s (\mu) 
\over \alpha_s (M_W)} \right)^{8/9} - \left( 
{\alpha_s (\mu) \over \alpha_s (M_W)} \right)^{-1/9} \right] \ \ ,
\label{sd8}
\eeqa
with 
\beq
\alpha_s (\mu) = \left[ 1 + 9 {\alpha_s (\mu) \over 4 \pi}
\ln \left( {M_W^2 \over \mu^2} \right) \right] \alpha_s (M_W) \ \ .
\label{sd9}
\eeq
An expansion of Eq.~(\ref{sd7}) through second order 
in $\alpha_s (\mu)$ gives 
\beqa
{\cal M} &\simeq& {G_F \over 2 \sqrt{2}F_\pi^2} 
\bigg[ \langle {\cal O}_1 \rangle_{\mu} \ 
+ \ { 3 \over 8 \pi} \ln \left( {M_W^2 \over \mu^2} 
\right) \langle \alpha_s {\cal O}_8 \rangle_{\mu} \ 
\nonumber \\
& &  \phantom{xxx} + {3\over 32 \pi^2} 
\ln^2 \left( {M_W^2 \over \mu^2} \right) \left( 
{8 \over 3} \langle \alpha_s^2 {\cal O}_1 \rangle_{\mu} 
- \langle \alpha_s^2 {\cal O}_8 \rangle_{\mu} \right) \bigg] \ \ .
\label{sd10}
\eeqa
Let us gain some feeling for the numbers involved.  
The minimum value of renormalization scale considered in this 
paper is $\mu_0 = 2$~GeV.  Adopting this scale and taking 
$\alpha_s (M_W) = 0.119$ and $\alpha_s (\mu_0) = 0.334$~\cite{aleph}, 
we find for the RG coefficients in Eq.~(\ref{sd7}), 
\beq
{\cal M} \simeq {G_F \over 2 \sqrt{2}F_\pi^2} 
\bigg[ 1.071 \langle {\cal O}_1 \rangle_{\mu_0} \ 
+ \ 0.268 \langle {\cal O}_8 \rangle_{\mu_0} 
\bigg] \ \ ,
\label{sd10a}
\eeq
whereas the coefficients in the perturbative expression 
of Eq.~(\ref{sd10}) become 
\beq
{\cal M} \simeq {G_F \over 2 \sqrt{2}F_\pi^2} 
\bigg[ \langle {\cal O}_1 \rangle_{\mu_0} \ 
+ \ 0.88 \langle \alpha_s {\cal O}_8 \rangle_{\mu_0} 
- 0.52 \left( \langle \alpha_s^2 {\cal O}_8 \rangle_{\mu_0} 
- {8 \over 3} \langle \alpha_s^2 {\cal O}_1 \rangle_{\mu_0} 
\right) \bigg] \ \ .
\label{sd10b}
\eeq

Finally, the condition of scale independence 
for the matrix element ${\cal M}$, 
\beq
\mu^2 {\partial \over \partial \mu^2} ~{\cal M} = 0 \ \ ,
\label{sd11}
\eeq
leads directly to the renormalization group equations 
\beqa
\mu^2 {\partial \over \partial \mu^2} 
\langle {\cal O}_1 \rangle_{\mu} &=& {3 \over 8 \pi} 
\langle \alpha_s {\cal O}_8 \rangle_{\mu} \ \ ,
\label{sd12} \\
\mu^2 {\partial \over \partial \mu^2} 
\langle \alpha_s {\cal O}_8 \rangle_{\mu} &=& 
{1 \over 4 \pi} \left[ {16 \over 3} \langle \alpha_s^2 
{\cal O}_1 \rangle_{\mu} - 2 \langle \alpha_s^2 
{\cal O}_8 \rangle_{\mu} \right] \ \ .
\label{sd13} 
\eeqa

To summarize --- the above operator-product analysis involves 
computing radiative corrections perturbatively to 
one-loop order in QCD ({\it cf} Eq.~(\ref{sd6})) 
and retaining only the dependence on leading logarithms 
in the evolution from scale $M_W$ down to scale $\mu$.  
The value of $\mu$ cannot be taken too small, otherwise 
the perturbative framework breaks down.  

\subsection{Verification of the Operator Product Expansion}
Despite the explicit difference between the procedures of Sect.~II 
and that carried out directly above, the two are 
equivalent.  In particular, we can show that
 the $\langle {\cal O}_1 \rangle_{\mu}$ 
sum rule of Eq.~(\ref{sp5}) and the $\langle {\cal O}_8 \rangle_{\mu}$ 
sum rule of Eq.~(\ref{sr31}) reproduce the OPE to the leading log level.  
This verifies both the derivations that we provided and gives a 
direct insight into the workings of the OPE.

Consider a partition of ${\cal M}$ 
characterized by the scale $\mu$, 
\beq
{\cal M} = {\cal M}_<(\mu) + {\cal M}_>(\mu) \ \ ,
\label{c6}
\eeq
where ${\cal M}_<(\mu)$ and ${\cal M}_>(\mu)$ are 
dependent respectively on contributions with 
$Q < \mu$ and $Q > \mu$.  Also, in addition to maintaining 
the requirement that $\mu$ lie in the pQCD domain, 
we further constrain it to obey $\mu \ll M_W$.  We then obtain 
\beqa
{\cal M}_< (\mu) &=&  {3 G_F M_W^2 \over 32 \sqrt{2}\pi^2 
F_\pi^2} \int_0^{\mu^2} dQ^2 \ {Q^4 \over Q^2 + M_W^2} 
\left[ \Pi_{V,3} (Q^2) - \Pi_{A,3} (Q^2) \right] 
\nonumber \\
&=& {3 G_F \over 32 \sqrt{2}\pi^2 
F_\pi^2} \int_0^{\mu^2} dQ^2 \ Q^4 
\left[ \Pi_{V,3} (Q^2) - \Pi_{A,3} (Q^2) \right] 
+ {\cal O}(\mu^2 / M_W^2) 
\label{va6} 
\eeqa
and 
\beq
{\cal M}_>(\mu) = {3 G_F M_W^2 \over 32 \sqrt{2}\pi^2 
F_\pi^2} \int_{\mu^2}^\infty dQ^2 \ {Q^4 \over Q^2 + M_W^2} 
\left[ \Pi_{V,3} (Q^2) - \Pi_{A,3} (Q^2) \right] \ \ .
\label{va7}
\eeq
Upon inserting the large-$Q$ form of Eq.~(\ref{c2}) into 
Eq.~(\ref{va7}), we obtain 
\beqa
{\cal M}_>(\mu) &=& { G_F \over  \sqrt{2} F_\pi^2} 
\bigg[ {3 \over 8 \pi} 
\ln \left( {M_W^2 \over \mu^2} \right) 
\langle \alpha_s {\cal O}_8 \rangle_{\mu} 
\nonumber \\
& & - {3 \over 32\pi^2} 
\ln^2 \left( {M_W^2 \over \mu^2} \right) 
\left( \langle \alpha_s^2 {\cal O}_8 \rangle_{\mu} 
- {8\over 3} \langle \alpha_s^2 {\cal O}_1 \rangle_{\mu} 
\right) \bigg] \ \ .
\label{va11}
\eeqa
Comparison of Eq.~(\ref{sd10}) with Eq.~(\ref{va11}) 
yields the relation, 
\beq
{\cal M}_<(\mu) = { G_F \over 2 \sqrt{2} F_\pi^2} 
\langle {\cal O}_1 \rangle_{\mu} \ \ .
\label{va12}
\eeq

We see that the operators that we originally defined independently of the weak
interaction are in fact the ones that appear in the Operator Product Expansion.
Of course, this is to be expected, but it provies an explicit pedagogical
demonstration of the nature of the OPE. 

\subsection{Sum Rules and RG Relations}
To complete the chain of logic, we demonstrate consistency of the 
spectral function sum rules for $\langle {\cal O}_1 
\rangle_{\mu}^{\rm (c.o.)}$ and 
$\langle {\cal O}_8 \rangle_{\mu}^{\rm (c.o.)}$ 
with the corresponding renormalization group relations obtained 
previously from the operator product expansion 
({\it cf} Eqs.~(\ref{sd12}),(\ref{sd13})).  The RG equation 
for $\langle {\cal O}_1 \rangle_{\mu}^{\rm (c.o.)}$ is immediately 
recovered upon differentiating the sum rule of Eq.~(\ref{sp5}) 
and making use of Eq.~(\ref{sr31}),
\beq
\mu^2 {\partial \over \partial \mu^2} \langle 
{\cal O}_1 \rangle_{\mu}^{\rm (c.o.)} = {3 \over 8 \pi} \cdot 
{\mu^2 \over 2 \pi} \int_0^\infty ds\  {s^2 \over s + \mu^2}~\left[ 
\rho_{V,3} - \rho_{A,3} \right](s) = {3 \over 8 \pi} 
\langle \alpha_s {\cal O}_8 \rangle_{\mu}^{\rm (c.o.)} \ \ .
\label{sp6}
\eeq

To derive the RG relation for $\langle {\cal O}_8 
\rangle_{\mu}^{\rm (c.o.)}$ we start with the sum rule of Eq.~(\ref{sr31}),
\beq
\mu^2 {\partial \over \partial \mu^2} \langle 
{\cal O}_8 \rangle_{\mu}^{\rm (c.o.)} = {\mu^2 \over 2 \pi }  
{\partial \over \partial \mu^2} 
\int_0^\infty ds\  {\mu^2 \over s + \mu^2}~ s^2 \left[ 
\rho_{V,3} - \rho_{A,3} \right](s) \ \ .
\label{sp7}
\eeq
The integral in the above is seen to be $\mu^6 \left(\Pi_{V,3} 
(\mu^2) - \Pi_{A,3} (\mu^2) \right)$.  We replace it using 
the aymptotic expression of Eq.~(\ref{sp2p}) to obtain 
\beqa
& & \mu^2 {\partial \over \partial \mu^2} \langle 
{\cal O}_8 \rangle_{\mu}^{\rm (c.o.)} = 
{\mu^2 \over 2 \pi }  {\partial \over \partial \mu^2} \Bigg[
\ln \left( {\mu^2 \over \Lambda^2} \right) ~
\left[ {8 \over 3} \langle \alpha_s^2 {\cal O}_1 
\rangle_{\mu}^{\rm (c.o.)} - \langle \alpha_s^2 
{\cal O}_8 \rangle_{\mu}^{\rm (c.o.)}  
\right] 
\nonumber \\
& & \phantom{xxxxxxxxxx} + \int_0^{\Lambda^2} ds\  s^2 ~\left[ 
\rho_{V,3} - \rho_{A,3} \right](s) + {\cal O}(\mu^{-2}) \Bigg]\ \ , 
\label{sp8}
\eeqa
from which the RG relation of Eq.~(\ref{sd13}) follows directly.

\section{${\overline {MS}}$ Renormalization} 
The work of the preceding sections was based on a 
momentum-space cutoff renormalization scheme, which 
is useful in yielding sum rules directly related to 
experimental data.  At the same time, however, it is 
distinct from the more standard ${\overline{MS}}$ prescription.  
In this Section, we demonstrate how to relate 
the two approaches.

\subsection{Short Distance Analysis} 
Let us reconsider Eq.~(\ref{r2}).  We can show how the cutoff 
renormalization is related to dimensional regularization by keeping 
the high-$Q^2$ part of the integral in Eq.~(\ref{r2}) and analyzing 
it in terms of an $\epsilon$-expansion. We divide the integral into 
integration ranges below and above $\mu^2$. For the part of 
integral with $Q^2$ below $\mu^2$, we can let $d \to 4$ and 
recover exactly the cutoff integrand of Eq.~(\ref{r3}), 
\beq
\langle {\cal O}_1 \rangle_{\mu}^{\rm (d.r.)} = 
\langle {\cal O}_1 \rangle_{\mu} + 
{ (d - 1) \mu^{4 -d} \over (4 \pi)^{d/2} 
\Gamma(d/2)} \int_{\mu^2}^\infty dQ^2 \ Q^d 
( \Pi_{V,3} - \Pi_{A,3} )(Q^2) \ \  
\eeq
The asymptotic tail can be analysed in $d$ dimensions. 
 This introduces scheme dependence 
depending on which method is used to define Dirac algebra 
away from four dimensions, {\it e.g.} the 
naive dimensional regularization (NDR) and 
t'Hooft-Veltman (HV) schemes in which $\gamma_5$ is respectively 
anticommuting and commuting.~\cite{evan}  We find
\beq
{1 \over 3}(d-1) Q^d\cdot ( \Pi_{V,3} - \Pi_{A,3} )(Q^2) = 2 \pi \alpha_s
\langle {\cal O}_8 \rangle_{\mu}~Q^{d - 6} 
\left[ 1 + \left(d_s + {1 \over 4}\right) \epsilon \right] 
+ {\cal O}(\alpha_s^2) \ \ .
\label{r5}
\eeq
where $\epsilon \equiv 4 - d$ and 
$d_s \equiv d_{\rm scheme}$ is associated 
with the loop integration and scheme-dependence.   The values of 
$d_s$ in the NDR and HV schemes are 
\beq
d_s = \left\{ \begin{array}{ll} -5/6 & ({\rm NDR}) \\
\ ~1/6 & ({\rm HV}) \ \ . \end{array}\right.
\label{mart}
\eeq
The $Q^2 >\mu^2$ integral can then be performed with the result
\beq
\langle {\cal O}_1 \rangle_{\mu}^{\rm (d.r.)} =
 \langle {\cal O}_1 \rangle_{\mu} + 
{3 \over 16 \pi^2} \left[ {2 \over 4 - d} - \gamma + \ln 4\pi +
{3\over 2} +2d_s\right] \langle \alpha_s {\cal O}_8 \rangle_{\mu} \ \ .
\label{r4}
\eeq
The ${\overline{MS}}$ prescription is a subcase of dimensional
regularization in which the terms $2/(4-d) - \gamma + \ln 4\pi$ 
in Eq.~(\ref{r4}) are removed in the renormalization procedure.
%
This gives our desired relation in a given scheme, 
\beq
\langle {\cal O}_1 \rangle_\mu^{\rm {\overline {MS}}}  = 
\langle {\cal O}_1 \rangle_\mu^{\rm (c.o.)}  
+ {3 \alpha_s \over 8 \pi} 
\left( {3 \over 2} + 2 d_s \right) 
\langle {\cal O}_8 \rangle_\mu \ \ .
\label{va37}
\eeq

To derive an analogous relation between 
$\langle {\cal O}_8 \rangle_\mu^{\rm {\overline {MS}}}$ and 
$\langle {\cal O}_8 \rangle_\mu^{\rm (c.o.)}$ 
we employ the leading behavior of 
correlators and spectral functions in the ${\rm {\overline {MS}}}$ 
renormalization prescription, which has been calculated using the 
NDR scheme,~\cite{lanin}
\beqa
& & \phantom{xxxxx} Q^6 (\Pi_{V,3} - \Pi_{A,3})(Q^2) \ \sim \ 
2 \pi \langle \alpha_s {\cal O}_8 
\rangle_{\mu}^{\rm ({\overline {MS}})} 
\label{va71} \\
& & + \left[ 2 + {8 \over 3} \ln \left( {Q^2 \over \mu^2} \right) 
\right] \langle \alpha_s^2 
{\cal O}_1 \rangle_{\mu}^{\rm ({\overline {MS}})}
+ \left[ {119\over 12} - \ln \left( {Q^2 \over \mu^2} \right) 
\right] \langle \alpha_s^2 {\cal O}_8 
\rangle_{\mu}^{\rm ({\overline {MS}})}
\nonumber 
\eeqa
and 
\beq
(\rho_{V,3} - \rho_{A,3})(s) \sim {1 \over  s^3} 
\left[ {8 \over 3} 
\langle \alpha_s^2 {\cal O}_1 
\rangle_\mu^{\rm ({\overline {MS}})} 
-  \langle \alpha_s^2 
{\cal O}_8 \rangle_\mu^{\rm ({\overline {MS}})} 
\right] + \dots \ \ .
\label{va8}
\eeq
Then by setting  $Q=\mu$ in Eq.~(\ref{va71}) and combining the result 
with Eq.~(\ref{va8}) we find
\beq
\mu^2 \int_0^\infty ds\  {s^2 \over s + \mu^2}~\left[ 
\rho_{V,3} - \rho_{A,3} \right](s) = 2 \pi \alpha_s 
\left[ \left( 1 + 
{119 \alpha_s \over 24\pi} \right) \langle {\cal O}_8 
\rangle_{\mu}^{\rm ({\overline {MS}})} 
+ {\alpha_s \over \pi} \langle 
{\cal O}_1 \rangle_{\mu}^{\rm ({\overline {MS}})} \right] 
\ \ .
\label{sr1}
\eeq
Comparison with Eq.~(\ref{va37}) then implies that the NDR matrix 
element is given to first 
order in $\alpha_s$ by
\beq
\langle {\cal O}_8 \rangle_{\mu}^{\rm ({\overline {MS}})} = 
\left( 1 - {119 \alpha_s \over 24\pi} \right) 
\langle {\cal O}_8 \rangle_{\mu}^{\rm (c.o.)} 
 - {\alpha_s \over \pi} 
\langle {\cal O}_1 \rangle_{\mu} \ \ .
\label{sr44}
\eeq

\subsection{${\rm {\overline {MS}}}$ Matching at One Loop}

\begin{figure}
\vskip .1cm
\hskip 2.0cm
\psfig{figure=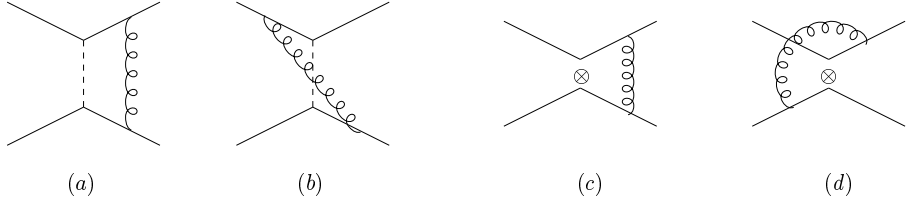,height=1.in}
\caption{Some QCD corrections: full theory (a)-(b), 
effective theory (c)-(d).\hfill 
\label{fig:drfig1}}
\end{figure}

In this section we perform the matching at one loop and
verify the scheme independence of the result. 
The effective operator ${\cal O}_{\Delta S = 1}^{\rm (eff)}$, 
is expressed in terms of
local operators ${\cal O}_1$ and ${\cal O}_8$, 
\beq
{\cal O}_{\Delta S = 1}^{\rm (eff)} = {G_F \over 2 \sqrt{2}} 
\left[ c_1 (\mu) {\cal O}_1   
+ c_8 (\mu){\cal O}_8  \right] \ \ . 
\label{sd71}
\eeq 
Determination of the coefficients $c_1$ and $c_8$ proceeds 
in two steps: first calculate QCD radiative corrections in 
both the full and effective theories, and then `match' the 
two calculations.  We shall carry out this procedure 
at one-loop order of the QCD radiative corrections. Since 
$c_1 = 1 + {\cal O}(\alpha_s^2)$ and $c_8 = {\cal O}(\alpha_s)$, 
this will yield a determination of $c_8$.  For 
definiteness, we consider the free scattering of zero momentum 
quarks and adopt a common quark mass $m$ to serve as the infrared cutoff. 

In the full theory, evaluations of the one-loop radiative 
corrections like those displayed in Figs.~\ref{fig:drfig1}(a)-(b) 
are finite and yield 
\beq
{\cal M} = {G_F \over 2 \sqrt{2} } \left[ 
\langle qq| {\cal O}_1 |qq\rangle_{\rm tree} 
+ {3 \alpha_s \over 8 \pi} 
\left( \ln {M_W^2 \over m^2} - 1 
\right) \langle qq|{\cal O}_8 |qq\rangle_{\rm tree} 
+ {\cal O}(\alpha_s^2) \right] \ \ .
\label{va33}
\eeq 
The analogous calculation in the effective theory 
is divergent and must be regularized.  We employ dimensional 
regularization which introduces the scheme dependence mentioned above. 
Our calculation of amplitudes like those in 
Figs.~\ref{fig:drfig1}(c)-(d) gives 
\beqa
& & {\cal M} = {G_F \over 2 \sqrt{2} } \bigg[ 
\langle qq|{\cal O}_1 |qq\rangle_{\rm tree} + {3 \alpha_s \over 8 \pi} 
\left( 1 + d_s  \epsilon + d_\ell \epsilon \right) 
\left[ {2 \over \epsilon} - \gamma + \ln 4\pi - \ln {m^2 \over \mu^2} 
\right] \langle qq|{\cal O}_8 |qq\rangle_{\rm tree} 
\nonumber \\
& & \phantom{xxxxxxxxxxx} 
+ c_8 \langle qq|{\cal O}_8 |qq\rangle_{\rm tree} \bigg] \ \ ,
\label{va34}
\eeqa 
After the removal of the divergent term in ${\overline {MS}}$ 
renormalization,
we compare the full theory and the effective theory 
to identify the coefficient function as 
\beq
c_8(\mu) = {3 \alpha_s \over 8\pi} \left[ \ln {M_W^2 \over \mu^2} 
- {3 \over 2} - 2 d_s \right] \ \ .
\eeq

When ${\cal O}_{\Delta S = 1}^{\rm (eff)}$ is applied to our 
problem of vacuum matrix elements, we have the amplitude
\beq
{\cal M} = {G_F \over 2 \sqrt{2} F_\pi^2} 
\left[ c_1 (\mu) \langle {\cal O}_1 \rangle_\mu^{\overline {MS}}   
+ c_8 (\mu) \langle {\cal O}_8 \rangle_\mu^{\overline {MS}} \right] \ \ . 
\label{sd71a}
\eeq 
Given our previous identification of the scheme dependent operator
$O_1$ in Eq.~(\ref{va37}), it can be seen that the scheme dependence cancels
between that of the matrix element and of the coefficient
in the operator product expansion.


\section{Numerical Analysis and Uncertainties}
We base our numerical determination of 
$\langle {\cal O}_8 \rangle_{\mu}^{\rm (c.o.)}$ and 
$\langle {\cal O}_1 \rangle_{\mu}^{\rm (c.o.)}$ respectively on the 
sum rules in Eq.~(\ref{sr31}) and Eq.~(\ref{sp5}).  
This involves calculation of the integrals 
$I_8$ and $I_1$, which contain the combination of spectral 
functions $\rho_{V,3} - \rho_{A,3}$,   
\beq
I_i = \int_0^\infty ds\  K_i (s,\mu^2) 
~\left[ \rho_{V,3} - \rho_{A,3} \right](s) \qquad (i = 1,8) \ \ ,
\label{f1a}
\eeq
with 
\beq
K_8 = s^2 {\mu^2 \over s + \mu^2} \ ,
\qquad 
K_1 \equiv s^2 \ln \left({s + \mu^2 \over s}\right) \ \ .
\label{f1b}
\eeq
As such, $I_8$ and $I_1$ belong to the 
family of spectral integrals which include the DMO sum rule~\cite{dmo},
the two Weinberg sum rules~\cite{sw} and the sum rule 
for the pion electromagnetic mass splitting~\cite{dgmly}.  The 
kernels occurring in these `classical' sum rules are 
\beq
K_{\rm DMO} = 1/s \ , \quad K_{\rm W1} = 1 \ , 
\quad K_{\rm W2} = s \ , \quad K_{\rm em} = s \ln~s \ \ .
\label{f2}
\eeq
This happenstance is most fortunate as the integrals defined 
by the kernels in Eq.~(\ref{f2}) form a powerful set of 
constraints for any evaluation of $I_8$ and $I_1$.  
Using an updated form of our earlier study~\cite{dg2} 
of chiral sum rules, we find for renormalization scale 
$\mu = 2$~GeV the values 
\beq
I_8 = - (0.30 \pm 0.04)\cdot 10^{-2} \ , 
\qquad 
I_1 = - (0.42 \pm 0.06)\cdot 10^{-2} \ , 
\label{f3a}
\eeq 
At the higher renormalization scale, $\mu = 4$~GeV, we obtain 
\beq
I_8 = - (0.43 \pm 0.06)\cdot 10^{-2} \ , 
\qquad 
I_1 = - (0.97 \pm 0.12)\cdot 10^{-2} \ .
\label{f3b}
\eeq 

\subsection{Uncertainties from Data Analysis}
The error bars quoted above correspond to our estimate of the 
uncertainty in the sum rules due to imprecision in our knowledge of 
the spectral functions. Before proceeding let us describe how these
were arrived at, and assess other sources of uncertainty. The data
at lower values of $s$ are extremely well known, and they introduce
very little uncertainty compared to other sources which we are 
concerned with. The high energy tail of $\rho_V -\rho_A$ is  
small above $s=5$~GeV$^2$. In the $\mu=2$~GeV integrals, there remains
essentially no sensitivity to this high energy tail once the constraints are
imposed\footnote{At higher values of $\mu$ there occurs more sensitivity to the
asymptotic tail, and it is the tail that describes the logarithmic running of
the $O_8$ matrix element. While we have a good handle on the size of the $1/s^3$
component of the tail, we know little about the $1/s^4$ component. However as
long as the $1/s^4$ portion is not much larger than the $1/s^3$ behavior for 
$s>5$~GeV$^2$, its effect is within our quoted error bars.}.	 
It is in the matching of these two regions that
one encounters the greatest uncertainties. Fortunately, the four integral
constraints described above are very stringent and allow us to 
limit the uncertainties in this region. We have used several methods
to construct spectral functions which match the data within error bars and
yet satisfy our sum rule constraints. These give variations in our 
integrals of under 6\%. In addition, we have considered
the situation where the asymptotic form of the spectral function is 
reached only on the average, with a damped oscillating term that 
provides deviations from the average. Since our sum rules are equivalent to 
transforming back to euclidiean $Q^2$, these oscillations give exponentially
suppressed effects at large $\mu^2$ once integrated. Again the constraints
are very powerful in further limiting this effect, and our studies lead 
us to increase the uncertainty in the fit to 10\% 
to account for this form of variation. 

We also must account for the fact that the data and the input into
the constraints
are measured in a world where $m_\pi^2$ is not zero, yet we are interested 
in the result in the chiral limit. This introduces corrections of 
order $m_\pi^2 / m_\rho^2$ which is of order 3\%. In fact, since we know 
some of the physics involved in passing to the chiral limit, we could attempt
to make a realistic correction for the extrapolation to the chiral
limit. However, since this would appear to introduce some model dependence
into our procedure, we prefer to simply include the uncertainty as
an error bar. In practice, the effect which has the most sensitivity
for our results is the constraint of the pion electromagnetic mass
difference, since the kernel $K_{\rm em}$ bears the greatest 
resemblance to $K_1$ and $K_8$.  Work on the pion and kaon
electromagnetic mass differences indicates that the quark mass corrections
are somewhat larger than average. Therefore, to be conservative
we triple the canonical error estimate, leading us to quote a
9\% uncertainty for the extrapolation to the chiral limit. We have
added this in quadrature to the statistical uncertainty to arrive at 
the error bars cited above.  

\subsection{Uncertainty from the Operator Product Expansion}
Finally, we need to address the fact that it has become common 
to cite matrix
elements at a scale $\mu = 2$~GeV, which is a rather low scale for
perturbative QCD to be fully in the asymptotic region. In fact, our 
method can be used for any $\mu$, and we can check if the asymptotic
QCD behavior is obtained. For example, the renormalization group 
equations relate the $\mu$-dependence to the magnitudes of the 
operator matrix elements.  Although one of the relations 
(Eqs.~(\ref{sd12}),(\ref{sp6})) is automatically satisfied, 
we explicitly showed 
above that the second holds only if $\mu$ is large enough, 
{\it i.e.} that it is well into the region where the asymptotic tail of the
spectral functions becomes applicable.  It is easy to see from the data
alone that this is not the case at $\mu = 2$~GeV. Another way to 
state the same result is that there remain power corrections in
the sum rule, although the renormalization group states that the running
with $\mu$ should be only logarithmic. We believe (because of 
the generality of our framework) that this issue must also be 
present in the lattice results, and we urge the evaluation of
lattice matrix elements at larger values of $\mu$.

We do see such non-asymptotic behavior in our results. Actually, for
${\cal O}_1$ our method of cutting off the high frequency modes
of the current-current product is in accord with Wilson's original 
idea of the definition of a scale-dependent matrix element. Therefore,
our sum rule for $\langle {\cal O}_1 \rangle_{\mu}^{\rm (c.o.)}$ 
can be treated as a {\it 
definition} of this amplitude at any scale $\mu$, even if that scale
is not yet asymptotic. For ${\cal O}_8$, however, there is some 
uncertainty as to an ideal definition in the non-asymptotic region.
For example, the RG relation of Eqs.~(\ref{sd12}),(\ref{sp6})   
requires that we use
exactly our definition, yet this only is foolproof if the RGEs 
are fully valid. Equivalently, if this matrix element is defined via 
the coefficient of $Q^{-6}$ in the vacuum polarization, there can 
be order $Q^{-8}$ corrections remaining if one works in the 
non-asymptotic region. We see evidence of such power corrections.
Moreover, attempting to discard the $Q^{-8}$ effects leads to a 
larger matrix element. 
At $\mu = 4$~GeV, the corrections are rather modest, in line with other 
uncertainties that we have described. However, at $\mu=2$~GeV, these
non-asymptotic effects represent a significant intrinsic uncertainty. 
{\it We have taken these into account by combining two evaluations, one 
obtained by evaluating the sum rule at $\mu=4$~GeV and using the 
RGE to transform down to
$\mu=2$~GeV, and the other by direct evaluation of the sum rule at
the lower scale. We average these two and assign the difference as 
an independent error bar.} The error bar is chosen such that
a one-sigma variation reproduces the full range between the two methods of
evaluation. We do this for both matrix elements. The quoted error bar 
at $\mu = 4$~GeV is scaled down from the $\mu=2$~GeV values by a 
factor of four, as appropriate for quadratic power corrections.

\subsection{Conversion to ${\overline {MS}}$ Renormalization}
We now transform to the ${\overline {MS}}$ matrix elements.
The results of our direct evaluation at $\mu=2$~GeV
 leads to the matrix elements 
 
\beqa
& & \langle {\cal O}_8 \rangle_{2~{\rm GeV}}^{\rm ({\overline {MS}})} = 
-(0.67 \pm 0.09)\cdot 10^{-3}~{\rm GeV}^6 \ , 
\nonumber \\
& & \langle {\cal O}_1 \rangle_{2~{\rm GeV}}^{\rm ({\overline {MS}})} = 
-(0.70 \pm 0.10)\cdot 10^{-4} ~{\rm GeV}^6\ \ , 
\label{f4a}
\eeqa
where we have taken $\alpha_s (2~{\rm GeV}) \simeq 0.334$. When we evaluate 
the integrals at $\mu=4$~GeV and use the RGE to rescale back to $\mu=2$~GeV, we
instead obtain

\beqa
& & \langle {\cal O}_8 \rangle_{2~{\rm GeV}}^{\rm ({\overline {MS}})} = 
-(1.29 \pm 0.15 )\cdot 10^{-3}~{\rm GeV}^6 \ , 
\nonumber \\
& & \langle {\cal O}_1 \rangle_{2~{\rm GeV}}^{\rm ({\overline {MS}})} = 
-(1.02  \pm 0.10 )\cdot 10^{-4} ~{\rm GeV}^6\ \ , 
\label{ff4}
\eeqa
which is a measure of the potential non-asymptotic corrections found at low
values of $\mu$. As explained above, this leads us to quote our result as
\beqa
& & \langle {\cal O}_8 \rangle_{2~{\rm GeV}}^{\rm ({\overline {MS}})} = 
-(0.98 \pm 0.13 \pm 0.23)\cdot 10^{-3}~{\rm GeV}^6 \ , 
\nonumber \\
& & \langle {\cal O}_1 \rangle_{2~{\rm GeV}}^{\rm ({\overline {MS}})} = 
-(0.86 \pm 0.10 \pm 0.16)\cdot 10^{-4} ~{\rm GeV}^6\ \ , 
\label{f4b}
\eeqa The first 
error bar corresponds to the uncertainty in the evaluation of the sum rule
whereas the second is the potential non-asymptotic uncertainty defined
above. Note that the two matrix elements differ by an order of magnitude. 
The related $K^0 \to \pi\pi$ matrix elements are then
\beqa
\langle (\pi\pi)_{I=2} | {\cal Q}_7^{(3/2)} 
| K^0\rangle_{2~{\rm GeV}} & = & 
(0.43\pm 0.05 \pm 0.08) ~{\rm GeV}^3 \ \ , 
\nonumber \\
\langle (\pi\pi)_{I=2} | {\cal Q}_8^{(3/2)} 
| K^0\rangle_{2~{\rm GeV}} & = & 
(2.58 \pm 0.37 \pm 0.47) ~{\rm GeV}^3 \ \ .
\eeqa   
In the NDR scheme with $\mu = 2~{\rm GeV}$ this translates into
the following B-factor determinations,  
\beqa
& & B_7^{(3/2)}[{\rm NDR},\mu = 2~{\rm GeV}] 
\left( {0.1~{\rm GeV} \over m_s + {\hat m}} \right)^2 
= 0.55 \pm 0.07 \pm 0.10 \ \ , 
\nonumber \\
& & B_8^{(3/2)}[{\rm NDR},\mu = 2~{\rm GeV}] 
\left( {0.1~{\rm GeV} \over m_s + {\hat m}} \right)^2 
= 1.11 \pm 0.16 \pm 0.23\ \ .
\label{fff5}
\eeqa
Note that the combination of B-factor and quark masses is 
`physical', appearing in the formula for $\epsilon'/\epsilon$. 
The comparison of these results with some lattice evaluations
is hampered by the fact that our evaluation is of the 
full matrix elements, while most lattice calculations are of the
B-factors directly~\cite{gbs}. 
If large values of quark masses are used, our
results are larger that other estimates, yet for the currently
favored smaller quark masses the results are not inconsistent.
There is one recent lattice evaluation which provides absolute matrix
elements which we can compare to. The Rome group\cite{Donini:1999nn}
quotes the matrix element in 
in the the quenched approximation using $K \to\pi$ matrix elements plus the 
chiral relation between $K \to\pi$ and $K\to\pi\pi$. When the meson 
masses are taken as the kaon mass they find, in the NDR scheme at 
$\mu=2$~GeV, 
\beqa
\langle (\pi\pi)_{I=2} | {\cal Q}_7^{(3/2)} 
| K^0\rangle_{2~{\rm GeV}} & = & 
 (0.22 \pm 0.04) ~{\rm GeV}^3 \ \ , 
\nonumber \\
\langle (\pi\pi)_{I=2} | {\cal Q}_8^{(3/2)} 
| K^0\rangle_{2~{\rm GeV}} & = & 
(1.02 \pm 0.10)~{\rm GeV}^3 \ \ .
\eeqa 
The quoted error does not include estimates of the effect
of quenching nor the extrapolation to the continuum limit.
Their results seem to be systematically smaller than ours.

Finally at $\mu=4$~GeV we have the vacuum matrix elements, 
\beqa
& & \langle {\cal O}_8 \rangle_{4~{\rm GeV}}^{\rm ({\overline {MS}})} = 
-(1.63 \pm 0.20 \pm 0.06)\cdot 10^{-3} ~{\rm GeV}^6\ , 
\nonumber \\
& & \langle {\cal O}_1 \rangle_{4~{\rm GeV}}^{\rm ({\overline {MS}})} = 
-(1.71 \pm 0.20 \pm 0.04)\cdot 10^{-4}~{\rm GeV}^6 \ \ .
\label{f44}
\eeqa
The corresponding $K \to \pi\pi$ matrix elements are
\beqa
\langle (\pi\pi)_{I=2} | {\cal Q}_7^{(3/2)} 
| K^0\rangle_{4~{\rm GeV}} & = & 
 (0.85 \pm 0.10  \pm 0.02)~{\rm GeV}^3 \\ \nonumber
\langle (\pi\pi)_{I=2} | {\cal Q}_8^{(3/2)} 
| K^0\rangle_{4~{\rm GeV}} & = & 
(4.34 \pm 0.56 \pm 0.15)~{\rm GeV}^3
\eeqa 
and for the B-factors we obtain, 
\beqa
& & B_7^{(3/2)}[{\rm NDR},\mu = 4~{\rm GeV}] 
\left( {0.1~{\rm GeV} \over m_s + {\hat m}} \right)^2 
= 1.10 \pm 0.13 \pm 0.03\ \ , 
\nonumber \\
& & B_8^{(3/2)}[{\rm NDR},\mu = 4~{\rm GeV}] 
\left( {0.1~{\rm GeV} \over m_s + {\hat m}} \right)^2 
= 1.87 \pm 0.25 \pm 0.07 \ \ .
\label{f54}
\eeqa
That $B_7^{(3/2)}$ has a large variation with $\mu$ 
is expected from 
the RGE of Eq.~(\ref{sd13}), given our previous result that the 
vacuum matrix element of ${\cal O}_8$ is much larger than that of
${\cal O}_1$.

\section{Concluding comments}

The method that we have described has the virtue of being a fully rigorous
framework. Moreover, the input data is largely taken from experiment, and
hence represents an evaluation that is model independent.
Besides the direct comparison with the results with lattice calculations, there
may also be other lessons in this calculation. Since in our method 
the matix elements are
evaluated by constructing the Euclidean vacuum polarization function, lattice
calculations may also be able to directly follow many of the steps in our
procedure, and thereby test their methods in more detail. By explicitly 
studying the product of currents at non-zero values of the spatial separation,
the matrix elements can be evaluated
without some of the operator mixing problems that occur on the lattice when
using local operators. Moreover, by studying vacuum matrix elements as well as
hadronic matrix elements, the chiral relations can be checked on the lattice.
Finally, we recall the lesson, discussed above, that power-law corrections
still appear to exist at $\mu=2$~GeV, especially in the $O_8$ matrix element.
This raises the concern that when one is working at such a low value of $\mu$,
there may be significant corrections even in lattice evaluations. Certainly, use
of $\mu < 1$~GeV, as occurs in many model dependent evaluations, appears
extremely dubious.

The values displayed above 
are based on working in the chiral limit of massless quarks.  
One must, however, add to these the chiral corrections.  
Work has begun on this important problem.~\cite{cdg}
Nevertheless, it is interesting to look at the phenomenological
consequences of the results reported in this paper.. While we 
cannot give a full evaluation of $\epsilon' /\epsilon$ 
because we have not evaluated the contribution of $B_6$, we can 
give the contribution arising from the electroweak penguin, 
\beq
\left({\epsilon' \over \epsilon}\right)_{B_8} 
= (-12 \pm 3) \cdot 10^{-4} \ \ .
\eeq
The effect of $B_6$ is expected to be positive, and needs to be 
almost three times larger than that of $B_8$ 
if the Standard Model is to account for the
experimental value.

\acknowledgments

This work was supported in part by the National
Science Foundation.  We thank Guido Martinelli 
for useful comments.

\eject

\end{document}